\documentclass[a4paper,fleqn,usenatbib]{mnras}

\usepackage[T1]{fontenc}
\usepackage{ae,aecompl}

\usepackage{graphicx}	
\usepackage{amsmath}	
\usepackage{amssymb}	

\title[Radiative lifetimes and oscillator strengths in \ion{Co}{ii}]{Experimental radiative lifetimes for highly excited states and calculated oscillator strengths for lines of astrophysical interest in singly ionized cobalt (Co II)}

\author[P. Quinet et al.]{
P. Quinet,$^{1,2}$\thanks{E-mail: Pascal.Quinet@umons.ac.be}
V. Fivet,$^{1}$
P. Palmeri,$^{1}$
L. Engstr\"om,$^{3}$
H. Hartman,$^{4,5}$
\newauthor
H. Lundberg,$^{3}$
and H. Nilsson$^{4}$
\\
$^{1}$Physique Atomique et Astrophysique, Universit\'e de Mons, B-7000 Mons, Belgium\\
$^{2}$IPNAS, Universit\'e de Li\`ege, Sart Tilman, B-4000 Li\`ege, Belgium\\
$^{3}$Department of Physics, Lund University, Box 118, SE-221 00 Lund, Sweden\\
$^{4}$Lund Observatory, Lund University, Box 43, SE-221 00 Lund, Sweden\\
$^{5}$Material Sciences and Applied Mathematics, Malm\"o University, 20506 Malm\"o, Sweden
}

\date{Accepted XXX. Received YYY; in original form ZZZ}

\pubyear{2016}

\begin{document}
\label{firstpage}
\pagerange{\pageref{firstpage}--\pageref{lastpage}}
\maketitle

\begin{abstract}
This work reports new experimental radiative lifetimes and calculated oscillator strengths for transitions of astrophysical interest in singly ionized cobalt. More precisely, nineteen radiative lifetimes in Co$^+$ have been measured with the time-resolved laser-induced fluorescence technique using one- and two-step excitations. Out of these, seven belonging to the high lying 3d$^7$($^4$F)4d configuration in the energy range 90697 -- 93738 cm$^{-1}$ are new, and the other twelve from the 3d$^7$($^4$F)4p configuration with energies between 45972 and 49328 cm$^{-1}$ are compared with previous measurements. In addition, a relativistic Hartree-Fock model including core-polarization effects has been employed to compute transition rates. Supported by the good agreement between theory and experiment for the lifetimes, new reliable transition probabilities and oscillator strengths have been deduced for 5080 \ion{Co}{ii} transitions in the spectral range 114 -- 8744 nm.
\end{abstract}




\section{Introduction}

The final stage of exothermal elemental production in stars is the iron-group elements. The even-$Z$ atoms are produced by consecutive capture of helium nuclei, and named $\alpha$-elements. The production of the odd-$Z$ elements is not as well constrained, and does not follow the abundance trends of the $\alpha$-elements, indicating non-common production sites. As a result of this nucleosynthesis in the interior of stars, the even-$Z$ nuclei such as Ca, Ti, Cr, and Fe have a higher cosmic abundance compared to the odd-$Z$ nuclei located in between. However, the astrophysical interest for the odd-$Z$ iron-group elements has increased in recent years. In the present project, we target atomic data for \ion{Co}{ii} (Z=27).
Cobalt is believed to be produced primarily in type II supernova, and also to a lesser extent in type Ia (Woosley \& Weaver 1995, Bravo \& Martinez-Pinedo 2012, Battistini \& Bensby 2015). Abundance determinations in stars serve as important tests of the stellar evolution and supernova explosion models (Pagel 2009). Furthermore, high-excitation spectral lines have additional diagnostic value, since they can be used to benchmark non local thermodynamical equilibrium (non-LTE) modelling of stellar atmospheres. Along with the development of 3D model atmospheres, a trustworthy non-LTE treatment is the current challenge for accurate stellar abundances. High-precision atomic data for selected lines is important for this development (Lind {\it et al} 2012).

In the case of cobalt, a big effort was recently made by Lawler {\it et al} (2015) to provide improved oscillator strengths for about 900 lines belonging to the first spectrum (\ion{Co}{i}). These were deduced from emission branching fractions measured from hollow cathode lamp spectra recorded using a Fourier transform spectrometer and a high-resolution echelle spectrograph combined with radiative lifetimes determined by the time-resolved laser-induced fluorescence (TR-LIF) technique.

Concerning the second spectrum (\ion{Co}{ii}), several papers have reported experimental determinations of radiative data. Lifetime measurements have been performed by Pinnington {\it et al} (1974) and S\o rensen (1979) using the beam-foil technique and by Salih {\it et al} (1985) and Mullman {\it et al} (1998) using TR-LIF. Experimental transition probabilities were reported for 41 spectral lines by Salih {\it et al} (1985), Crespo Lopez-Urrutia {\it et al} (1994) and Mullman {\it et al} (1998) by combining branching fraction measurements with available lifetimes. Theoretical data have been published by Raassen {\it et al} (1998) who computed oscillator strengths for a large number of \ion{Co}{ii} lines using the method of orthogonal operators. However, all these studies were limited to radiative decays from the odd-parity 3d$^7$4p configuration to lower states belonging to the 3d$^8$ and 3d$^7$4s even-parity configurations. More extensive calculations are reported by Kurucz (2011).

The main goal of the present work is to extend the knowledge of radiative data to higher energy states and to provide a consistent set of transition rates for a large number of spectral lines in singly ionized cobalt. More precisely, new experimental lifetime measurements were performed by TR-LIF for 12 energy levels belonging to the 3d$^7$4p odd-parity configuration and 7 energy levels belonging to the 3d$^7$4d even-parity configuration using one- and two-step excitations, respectively. In addition, transition probabilities and oscillator strengths were computed for 5080 \ion{Co}{ii} lines in a wide spectral region, from ultraviolet to infrared, using a pseudo-relativistic Hartree-Fock model including core-polarization effects.

\section{Radiative lifetime measurements}

The experimental set-up for one- and two-step experiments at the Lund High Power Laser Facility has recently been described in detail (Engstr\"om {\it et al} 2014, Lundberg {\it et al} 2016). For an overview we refer to Figure 1 in Lundberg {\it et al} (2016), and here we mainly give the most important details.  The Co$^+$ ions were produced by focusing 10 ns long pulses from a frequency doubled Nd:YAG laser onto a rotating Co target placed in a vacuum chamber where the pressure was about 10$^{-4}$ mbar. The Co plasma created in the ablation process was crossed by one or two laser beams 5 mm above the target.

Both laser systems consisted of a frequency doubled Nd:YAG laser (Continuum NY-82) pumping a Continuum Nd - 60 dye laser, mostly operating with DCM dye. For the wavelengths above 225 nm the second (or only) laser used Oxazin dye instead. In all measurements the final output from the dye lasers was frequency tripled using KDP and BBO crystals. The pulse length from the first laser was 10 ns, whereas for the second (or only) laser we achieved a pulse length after tripling of about 1 ns by injection seeding and compressing the output from the Nd:YAG laser. The compressor utilized stimulated Brillouin scattering in water.

The fluorescence emitted by the Co$^+$ ions was dispersed by a 1/8 m grating monochromator, with its 0.1 mm wide entrance slit oriented parallel to the excitation laser beam, registered by a fast micro-channel-plate photomultiplier tube (Hamamatsu R3809U) and finally digitized by a Tektronix DPO 7254 oscilloscope. The oscilloscope, with 2.5 GHz analog band width, sampled the decay in 50 ps intervals. All measurements used the second spectral order, giving an observed line width of about 0.5 nm. The excitation laser pulse shape was recorded simultaneously using a fast photo diode and digitized in a second channel on the oscilloscope. The final decay curves and pulse shape were obtained by averaging over 1000 laser pulses.  The code DECFIT (Palmeri {\it et al} 2008) was then used to extract the lifetimes by fitting a single exponential function convoluted by the measured shape of the second-step laser pulse and a background function to the observed decay.

Table 1 gives the wavelengths for the excitation and detection channels used in the single-step measurements. All two-step excitations started from the 3d$^7$4p z$^5$G$_5$ level at 47346 cm$^{-1}$, and the excitation and detection channels used are listed in Table 2.

\begin{table*}
    \centering
    \caption{Single-step measurements for 3d$^7$($^4$F)4p levels in \ion{Co}{ii}.}
         \label{KapSou}
         $$
         \begin{tabular}{cccccccccc}
            \hline
            Level     & Energy$^a$ & Starting level$^a$ & Excitation$^b$ & Detection$^c$ \\
                        & E (cm$^{-1}$) & E (cm$^{-1}$)  & $\lambda$$_{air}$ (nm) & $\lambda$$_{air}$ (nm) \\
            \hline
            z $^5$F$_3^{\circ}$ & 45972.036 &    0.000 & 217.45 & 238, 241 \\
            z $^5$D$_4^{\circ}$ & 46320.832 &    0.000 & 215.82 & 232, 236 \\
            z $^5$F$_2^{\circ}$ & 46452.700 & 1597.197 & 222.86 & 239      \\
            z $^5$D$_3^{\circ}$ & 47039.105 &    0.000 & 212.52 & 232, 235 \\
            z $^5$G$_6^{\circ}$ & 47078.494 & 3350.494 & 228.62 & 229$^d$    \\
            z $^5$G$_5^{\circ}$ & 47345.845 &    0.000 & 211.15 & 231      \\
            z $^5$D$_2^{\circ}$ & 47537.365 &  950.324 & 214.58 & 233, 235 \\
            z $^5$G$_4^{\circ}$ & 47807.493 &  950.324 & 213.35 & 228, 231, 269 \\
            z $^5$G$_3^{\circ}$ & 48150.940 &  950.324 & 211.79 & 231      \\
            z $^5$G$_2^{\circ}$ & 48388.442 & 1597.197 & 213.65 & 231      \\
            z $^3$G$_4^{\circ}$ & 49348.304 &  950.324 & 206.55 & 253      \\
            \hline
         \end{tabular}
         $$
\begin{list}{}{}
\item[$^{\mathrm{a}}$] NIST compilation (Kramida {\it et al} 2014)
\item[$^{\mathrm{b}}$] All levels were excited using the third harmonic of the dye laser
\item[$^{\mathrm{c}}$] All fluorescence measurements were performed in the second spectral order
\item[$^{\mathrm{d}}$] Corrected for scattered laser light at the same wavelength
\end{list}
   \end{table*}

   \begin{table*}
   \centering
      \caption{Two-step measurements for 3d$^7$($^4$F)4d levels in \ion{Co}{ii}.}
         \label{KapSou}
         $$
         \begin{tabular}{cccccccccc}
            \hline
            Level     & Energy$^a$ & Excitation$^b$ & Detection$^c$ \\
                        & E (cm$^{-1}$) & $\lambda$$_{air}$ (nm) & $\lambda$$_{air}$ (nm) \\
            \hline
            e $^5$G$_5$ & 90697.464 & 230.60 & 220   \\
            e $^5$H$_6$ & 90975.540 & 229.13 & 236$^d$ \\
            e $^5$G$_4$ & 91049.431 & 228.74 & 221, 223 \\
            e $^3$G$_5$ & 91326.932 & 227.37 & 222   \\
            e $^3$H$_6$ & 91623.455 & 225.78 & 232$^d$ \\
            e $^5$H$_5$ & 91646.447 & 225.66 & 228$^d$ \\
            f $^3$F$_4$ & 93738.788 & 215.48 & 237$^d$ \\
            \hline
         \end{tabular}
         $$
\begin{list}{}{}
\item[$^{\mathrm{a}}$] NIST compilation (Kramida {\it et al} 2014)
\item[$^{\mathrm{b}}$] All levels were excited from the intermediate 4p z $^5$G$_5$ level at 47346 cm$^{-1}$, given in Table 1, using the third harmonic of the dye laser
\item[$^{\mathrm{c}}$] All fluorescence measurements were performed in the second spectral order
\item[$^{\mathrm{d}}$] Corrected for the strong fluorescence background from the intermediate level at 231 nm, as discussed in the text and illustrated in Figure 1
\end{list}
   \end{table*}

For the two-step measurements several special experimental aspects have to be considered. Before every measurement the delay between the two lasers was adjusted so that the short pulse from second laser coincided with the maximum population of the intermediate
3d$^7$4p z$^5$G$_5$ level. The latter was found by observing the maximum in the fluorescence light from this level at either 231 or 266 nm.  When detecting the decay of the final 3d$^7$4d levels two types of line blending can occur. The very intense light emitted by the intermediate level at 231 nm results in a time dependent background observable in a several nm wide wavelength region. The lifetime of z$^5$G$_5$ is about 3 ns but the fluorescence signal extends over more than 10 ns due to the width of the first-step laser pulse. This blending problem can, however, be accurately handled by recording the background signal with the second-step laser turned off and then subtracting this from the observed final decay. This technique is illustrated in Figure 1, and the measurements influenced by this effect are marked in Table 2. A more difficult problem is that the level under investigation usually decays along several cascade chains where one or more secondary (4s - 4p) channels may occur close to the measured decay. Contrary to blending from the intermediate level the cascades are of much lower intensity, and hence the blending has to be almost perfect in wavelength to cause a serious problem. Since such blends cannot be compensated for, we carefully investigated this possibility before choosing the appropriate channels to use, as discussed in Lundberg {\it et al} (2016).

The final lifetimes are given in Table 3 and represent the averages of 10 - 20 measurements performed over several days. The quoted uncertainties are mainly based on the variation between the different measurements (Palmeri {\it et al} 2008).

   \begin{table*}
   \centering
      \caption{Experimental and calculated radiative lifetimes (in ns) for selected energy levels belonging to the 3d$^7$4p and 3d$^7$4d configurations of \ion{Co}{ii}.}
         \label{KapSou}
         $$
         \begin{tabular}{cccccccccc}
            \hline
            Level     & Energy$^a$    & \multicolumn{2}{c}{Experiment} & \multicolumn{2}{c}{Calculations} \\
                        & (cm$^{-1}$) & Previous & This work & This work & Kurucz$^d$ \\
            \hline
            4p~ z $^5$F$_5$ & 45197.711 & 3.5 $\pm$ 0.2$^b$   &                 & 2.97 & 2.79 \\
            4p~ z $^5$F$_4$ & 45378.754 & 3.5 $\pm$ 0.2$^b$   &                 & 3.07 & 2.87 \\
            4p~ z $^5$F$_3$ & 45972.036 & 3.7 $\pm$ 0.2$^b$   & 3.21 $\pm$ 0.15 & 3.03 & 2.85 \\
            4p~ z $^5$D$_4$ & 46320.832 & 3.3 $\pm$ 0.2$^b$   & 3.10 $\pm$ 0.15 & 3.02 & 2.82 \\
            4p~ z $^5$F$_2$ & 46452.700 & 3.0 $\pm$ 0.2$^b$   & 3.12 $\pm$ 0.15 & 3.00 & 2.83 \\
            4p~ z $^5$F$_1$ & 46786.409 & 3.4 $\pm$ 0.2$^b$   &                 & 2.99 & 2.83 \\
            4p~ z $^5$D$_3$ & 47039.105 & 3.4 $\pm$ 0.2$^b$   & 3.11 $\pm$ 0.15 & 3.03 & 2.82 \\
            4p~ z $^5$G$_6$ & 47078.494 & 3.0 $\pm$ 0.3$^b$   & 2.70 $\pm$ 0.15 & 2.54 & 2.48 \\
            4p~ z $^5$G$_5$ & 47345.845 & 3.2 $\pm$ 0.2$^b$   & 2.93 $\pm$ 0.15 & 2.72 & 2.65 \\
            4p~ z $^5$D$_2$ & 47537.365 & 3.3 $\pm$ 0.2$^b$   & 3.04 $\pm$ 0.20 & 3.03 & 2.83 \\
            4p~ z $^5$G$_4$ & 47807.493 & 3.0 $\pm$ 0.2$^b$   & 2.84 $\pm$ 0.15 & 2.67 & 2.60 \\
            4p~ z $^5$D$_1$ & 47848.781 & 3.4 $\pm$ 0.2$^b$   &                 & 3.05 & 2.84 \\
            4p~ z $^5$G$_3$ & 48150.940 & 3.1 $\pm$ 0.2$^b$   & 2.86 $\pm$ 0.15 & 2.65 & 2.58 \\
            4p~ z $^5$G$_2$ & 48388.442 & 3.2 $\pm$ 0.2$^b$   & 2.84 $\pm$ 0.15 & 2.63 & 2.57 \\
            4p~ z $^5$G$_5$ & 48556.052 & 3.6 $\pm$ 0.2$^c$   & 2.84 $\pm$ 0.15 & 2.63 & 3.07 \\
            4p~ z $^3$G$_4$ & 49348.304 & 3.3 $\pm$ 0.2$^c$   & 3.03 $\pm$ 0.15 & 2.78 & 2.58 \\
            4p~ z $^3$F$_4$ & 49697.683 & 2.9 $\pm$ 0.2$^c$   &                 & 2.54 & 2.46 \\
            4p~ z $^3$G$_3$ & 50036.348 & 3.5 $\pm$ 0.2$^c$   &                 & 3.01 & 2.79 \\
            4p~ z $^3$F$_3$ & 50381.724 & 2.9 $\pm$ 0.2$^c$   &                 & 2.44 & 2.35 \\
            4p~ z $^3$F$_2$ & 50914.325 & 2.9 $\pm$ 0.2$^c$   &                 & 2.28 & 2.13 \\
            4p~ z $^3$D$_3$ & 51512.268 & 2.3 $\pm$ 0.2$^c$   &                 & 1.94 & 1.83 \\
            4p~ z $^3$D$_2$ & 52229.725 & 2.3 $\pm$ 0.2$^c$   &                 & 1.97 & 1.86 \\
            4p~ z $^3$D$_1$ & 52684.634 & 2.4 $\pm$ 0.2$^c$   &                 & 1.98 & 1.87 \\
            4d~ e $^5$G$_5$ & 90697.464 &                     & 1.41 $\pm$ 0.10 & 1.44 & 1.37 \\
            4d~ e $^5$H$_6$ & 90975.540 &                     & 1.57 $\pm$ 0.10 & 1.42 & 1.36 \\
            4d~ e $^5$G$_4$ & 91049.431 &                     & 1.35 $\pm$ 0.10 & 1.38 & 1.35 \\
            4d~ e $^3$G$_5$ & 91326.932 &                     & 1.44 $\pm$ 0.10 & 1.43 & 1.41 \\
            4d~ e $^3$H$_6$ & 91623.455 &                     & 1.36 $\pm$ 0.10 & 1.44 & 1.39 \\
            4d~ e $^5$H$_5$ & 91646.447 &                     & 1.52 $\pm$ 0.10 & 1.40 & 1.35 \\
            4d~ f $^3$F$_4$ & 93738.788 &                     & 1.58 $\pm$ 0.10 & 1.56 & 1.64 \\
            \hline
         \end{tabular}
         $$
\begin{list}{}{}
\item[$^{\mathrm{a}}$] NIST compilation (Kramida {\it et al} 2014)
\item[$^{\mathrm{b}}$] Salih {\it et al} (1985)
\item[$^{\mathrm{c}}$] Mullman {\it et al} (1998)
\item[$^{\mathrm{d}}$] Kurucz (2011)
\end{list}
   \end{table*}

\section{Oscillator strength calculations}

 To model the atomic structure and to compute the radiative decay rates in singly ionized cobalt, we considered a relativistic Hartree-Fock (HFR) model using the suite of computer codes originally developed by Cowan (1981), and subsequently modified to take core-polarization into account, giving rise to the HFR+CPOL approach (see e.g. Quinet {\it et al} 1999, 2002). In the present case, the configurations explicitly included in the configuration interaction expansions were the following : 3d$^8$ + 3d$^7$4s + 3d$^7$5s + 3d$^7$4d + 3d$^7$5d + 3d$^6$4s$^2$ + 3d$^6$4s5s + 3d$^6$4s4d + 3d$^6$4s5d + 3d$^5$4s$^2$4d + 3d$^5$4s$^2$5s  (even parity) and 3d$^7$4p + 3d$^7$5p + 3d$^7$4f + 3d$^7$5f + 3d$^6$4s4p + 3d$^6$4s5p + 3d$^6$4s4f + 3d$^6$4s5f + 3d$^5$4s$^2$4p + 3d$^5$4s$^2$4f (odd parity). The ionic core considered for the core-polarization model potential
and the correction to the dipole operator was a vanadium-like core, i.e. a 3d$^5$ Co$^{4+}$ core. The dipole polarizability for such a core is 1.01 a$_0^3$ according to Fraga {\it et al.} (1976). For the cut-off radius, we used the HFR mean value of the outermost 3d core orbital, i.e. 1.00 a$_0$.

Some radial integrals, considered as free parameters, were then adjusted with a
least-squares optimization program minimizing the discrepancies between the calculated
Hamiltonian eigenvalues and the experimental energy levels from the NIST database (Kramida {\it et al} 2014). In this compilation, \ion{Co}{ii} energy levels from Pickering {\it et al} (1998) and from Pickering (1998) were increased by 6.7 parts in 10$^8$ to account for the calibration correction found by Nave \& Sansonetti (2011). Moreover, the additive constant '+x' introduced by Pickering {\it et al} (1998) for the 3d$^7$($^2$F)4s and 3d$^7$($^2$F)4p energy levels was eliminated by adjusting these levels to fit the wavelengths of 20 observed intersystem spectral lines. Four of these lines were measured by Pickering {\it et al} (1998) while the other 16 lines were observed by Iglesias (1979). For the 3d$^8$, 3d$^7$4s, 3d$^7$5s and 3d$^7$4d even-parity configurations, the average energies ($E_{av}$), the electrostatic direct ($F^k$) and exchange ($G^k$) integrals, the spin-orbit ($\zeta$$_{nl}$), and the effective interaction ($\alpha$, $\beta$) parameters were allowed to vary during the fitting process. In addition, the average energies for the 3d$^7$5d and 3d$^6$4s$^2$ configurations were also included in the adjustment. For the few experimental even levels located above 111000 cm$^{-1}$ it was very difficult to establish an unambiguous correspondence with the calculated values so these levels were omitted in the semi-empirical process. In the odd parity configurations, only experimental levels located below 90000 cm$^{-1}$ were used to optimize some radial parameters in the 3d$^7$4p and 3d$^6$4s4p configurations. The fitting process was not applied to the highly excited levels reported in the NIST compilation as belonging to 3d$^7$5p and 3d$^7$4f configurations because most of them appeared to be very strongly mixed with experimentally unknown levels. For both parities, all the other radial electrostatic interaction parameters were fixed at 80\% of their {\it ab initio} HFR values.

The numerical values of the parameters adopted in the present calculations are reported in Tables 4 and 5 for even-parity and odd-parity configurations, respectively. This semi-empirical process led to average deviations with experimental energy levels equal to 110 cm$^{-1}$ (158 levels, 30 fitted parameters) for the even parity, and 127 cm$^{-1}$ (114 levels, 12 fitted parameters) for the odd parity.

      \begin{table*}
      \centering
      \caption{Radial parameters adopted in the HFR+CPOL calculations for the 3d$^8$, 3d$^7$4s, 3d$^7$5s and 4d$^7$4d even-parity configurations of \ion{Co}{ii}.}
         \label{KapSou}
     $$
         \begin{tabular}{llrrrl}
            \hline
            Config.     & Parameter & $Ab~initio$ & Fitted & Ratio & Note$^a$ \\
                          &           & (cm$^{-1}$) & (cm$^{-1}$) &      & \\
            \hline
            3d$^8$        & E$_{av}$       & 20895 & 11615 &        & \\
                          & F$^2$(3d,3d)   & 84335 & 70521 & 0.836  & \\
                          & F$^4$(3d,3d)   & 52008 & 43842 & 0.843  & \\
                          & $\alpha$       &     0 &    69 &        & \\
                          & $\beta$        &     0 &   294 &        & \\
                          & $\zeta$$_{3d}$ &   474 &   470 & 0.992  & \\[0.2cm]
            3d$^7$4s      & E$_{av}$       & 26351 & 27378 &        & \\
                          & F$^2$(3d,3d)   & 92176 & 79060 & 0.858  & \\
                          & F$^4$(3d,3d)   & 57195 & 51853 & 0.907  & \\
                          & $\alpha$       &     0 &    92 &        & \\
                          & $\beta$        &     0 & -1115 &        & \\
                          & $\zeta$$_{3d}$ &   526 &   508 & 0.967  & \\
                          & G$^2$(3d,4s)   & 10203 &  7908 & 0.775  & \\[0.2cm]
            3d$^7$5s      & E$_{av}$       &103359 & 105208&        & \\
                          & F$^2$(3d,3d)   & 93461 & 78524 & 0.840  & \\
                          & F$^4$(3d,3d)   & 58056 & 52676 & 0.907  & \\
                          & $\alpha$       &     0 & 74    &        & \\
                          & $\beta$        &     0 & -1053 &        & \\
                          & $\zeta$$_{3d}$ &   531 &  479  & 0.902  & \\
                          & G$^2$(3d,5s)   &  1846 &  1654 & 0.896  & \\[0.2cm]
            3d$^7$4d      & E$_{av}$       &108802 &111591 &        & \\
                          & F$^2$(3d,3d)   & 93577 & 77104 & 0.824  & \\
                          & F$^4$(3d,3d)   & 58134 & 48229 & 0.830  & \\
                          & $\alpha$       &     0 & 93    &        & \\
                          & $\beta$        &     0 & -5    &        & \\
                          & $\zeta$$_{3d}$ &   532 &  556  & 1.046  & \\
                          & $\zeta$$_{4d}$ &    13 &   13  & 1.000  & F \\
                          & F$^2$(3d,4d)   &  4680 &  3850 & 0.823  & R1 \\
                          & F$^4$(3d,4d)   &  1705 &  1403 & 0.823  & R1 \\
                          & G$^0$(3d,4d)   &  1924 &  1103 & 0.573  & R2 \\
                          & G$^2$(3d,4d)   &  1635 &   938 & 0.573  & R2 \\
                          & G$^4$(3d,4d)   &  1139 &   654 & 0.573  & R2 \\
            \hline
         \end{tabular}
      $$
\begin{list}{}{}
\item[$^{\mathrm{a}}$] F : fixed parameter value; Rn : fixed ratio between these parameters.
\end{list}
   \end{table*}

      \begin{table*}
      \centering
      \caption{Radial parameters adopted in the HFR+CPOL calculations for the 3d$^7$4p and 3d$^6$4s4p odd-parity configurations of \ion{Co}{ii}.}
         \label{KapSou}
     $$
         \begin{tabular}{llrrrl}
            \hline
            Config.     & Parameter & Ab~initio & Fitted & Ratio & Note$^a$ \\
                          &           & (cm$^{-1}$) & (cm$^{-1}$) &      & \\
            \hline
            3d$^7$4p      & E$_{av}$          & 66556 & 68604 &        & \\
                          & F$^2$(3d,3d)      & 92841 & 78444 & 0.845  & \\
                          & F$^4$(3d,3d)      & 57642 & 52043 & 0.903  & \\
                          & $\alpha$          &     0 &    89 &        & \\
                          & $\beta$           &     0 & -1001 &        & \\
                          & $\zeta$$_{3d}$    &   529 &   519 & 0.981  & \\
                          & $\zeta$$_{4p}$    &   364 &   438 & 1.203  & \\
                          & F$^2$(3d,4p)      & 14303 & 12883 & 0.901  & \\
                          & G$^1$(3d,4p)      &  5670 &  4830 & 0.852  & \\
                          & G$^3$(3d,4p)      &  4579 &  3205 & 0.700  & \\[0.2cm]
            3d$^6$4s4p    & E$_{av}$          &109002 &117652 &        & \\
                          & F$^2$(3d,3d)      &100034 & 80027 & 0.800  & F  \\
                          & F$^4$(3d,3d)      & 62434 & 49947 & 0.800  & F  \\
                          & $\alpha$          &     0 &  0    &        & F  \\
                          & $\beta$           &     0 &  0    &        & F  \\
                          & $\zeta$$_{3d}$    &   583 &  583  & 1.000  & F  \\
                          & $\zeta$$_{4p}$    &   487 &  487  & 1.000  & F \\
                          & F$^2$(3d,4p)      & 16205 & 14190 & 0.876  & R  \\
                          & G$^2$(3d,4s)      & 10171 &  8906 & 0.876  & R  \\
                          & G$^1$(3d,4p)      &  5922 &  5186 & 0.876  & R  \\
                          & G$^3$(3d,4p)      &  5025 &  4400 & 0.876  & R  \\
                          & G$^1$(4s,4p)      & 45716 & 40030 & 0.876  & R  \\
            \hline
         \end{tabular}
      $$
\begin{list}{}{}
\item[$^{\mathrm{a}}$] F : fixed parameter value; R : fixed ratio between these parameters.
\end{list}
   \end{table*}

\section{Results and discussion}

The computed radiative lifetimes obtained in the present work are compared with the available experimental
values in Table 3. Our experimental values agree within the mutual error bars with Salih {\it et al} (1985) and Mullman {\it et al} (1998). However, there is a tendency for our new values to be somewhat shorter. This could probably be explained by our shorter pulse length and better time resolution in the detection system. As shown in this table, the overall agreement between theory and experiment is
good, the average relative differences being found to be equal to 12\%, 17\% and 5\%, when considering the lifetime measurements due to Salih {\it et al} (1985), Mullman {\it et al} (1998) and the present study, respectively. For comparison, Table 3 includes the theoretical lifetimes obtained by Kurucz (2011). This work also used a semi-empirical approach based on a superposition of configurations calculation with a modified version of the Cowan (1981) codes and experimental level energies. We note a very good qualitative agreement between the two calculations, but that our new results, with two exceptions, are consistently 5\% longer and closer to the experimental values.

Table 6 gives the HFR+CPOL oscillator strengths (log $gf$) and weighted transition probabilities ($gA$), together with the experimental energies of the lower and upper levels and the corresponding wavelengths for 5080 \ion{Co}{ii} spectral lines from 114 to 8744 nm. These correspond to all the electric dipole transitions involving the experimentally known levels (Pickering 1998, Pickering {\it et al} 1998, Kramida {\it et al} 2014) for which log $gf$ is larger than -4. In the last column of the table, we also give the value of the cancellation factor ($CF$), as defined by Cowan (1981). Very small values of this factor (typically $<$ 0.05) indicate strong cancellation effects in the calculation of the line strengths and the corresponding transition rates can be affected by larger uncertainties and should be considered with some care. This concerns about 25\% of the lines listed in Table 6, the large majority of them being characterized by weak oscillator strengths (log $gf$ $<$ -2).

\begin{table*}
\centering
\caption{Transition probabilities and oscillator strengths for Co II lines. XE+Y stands
for X $\times$ 10$^Y$. Only transitions between experimentally known energy levels with log $gf$ $\ge$ -4.0 are listed in the table. The full table is available online.}
\label{Kapsou}
$$
\begin{tabular}{rrccrccrrr}
\hline
$\lambda$$^a$ (nm)  & \multicolumn{3}{c}{Lower level$^b$}        & \multicolumn{3}{c}{Upper level$^b$} & \multicolumn{3}{c}{HFR+CPOL$^c$}  \\
                    & $E$ (cm$^{-1}$) & Parity & $J$                 & $E$ (cm$^{-1}$) & Parity & $J$          & log $gf$ & $gA$ (s$^{-1}$) & $CF$ \\
\hline
114.794	&	0	&	(e)	&	4	&	87112	&	(o)	&	4	&	-3.04	&	4.64E+06	&	0.058	\\
115.014	&	950	&	(e)	&	3	&	87896	&	(o)	&	3	&	-3.20	&	3.20E+06	&	0.051	\\
115.180	&	1597	&	(e)	&	2	&	88418	&	(o)	&	2	&	-3.30	&	2.53E+06	&	0.056	\\
116.231	&	1597	&	(e)	&	2	&	87633	&	(o)	&	3	&	-3.78	&	8.16E+05	&	0.027	\\
117.587	&	0	&	(e)	&	4	&	85044	&	(o)	&	3	&	-3.50	&	1.51E+06	&	0.171	\\
118.741	&	0	&	(e)	&	4	&	84217	&	(o)	&	3	&	-1.01	&	4.61E+08	&	0.316	\\
118.849	&	0	&	(e)	&	4	&	84141	&	(o)	&	4	&	-1.63	&	1.10E+08	&	0.098	\\
119.242	&	0	&	(e)	&	4	&	83863	&	(o)	&	3	&	-2.94	&	5.32E+06	&	0.016	\\
119.386	&	3350	&	(e)	&	5	&	87112	&	(o)	&	4	&	-3.89	&	6.09E+05	&	0.039	\\
119.841	&	950	&	(e)	&	3	&	84395	&	(o)	&	2	&	-1.18	&	3.07E+08	&	0.359	\\
...     & ...   & ...   & ...   & ...       & ...   & ...   & ...       &  ...          & ...       \\
\hline
\end{tabular}
$$
\begin{list}{}{}
\item[$^{\mathrm{a}}$] Vacuum ($\lambda$ $<$ 200 nm) and air ($\lambda$ $>$ 200 nm) Ritz wavelengths deduced from the experimental energy level values compiled at NIST (Kramida {\it et al} 2014).
\item[$^{\mathrm{b}}$] From the NIST database (Kramida {\it et al} 2014). For commodity, energy values have been rounded to the nearest unit in the table; for more accurate values, see the NIST compilation. \\
\item[$^{\mathrm{c}}$] This work.
\end{list}
\end{table*}

When comparing the results obtained with our computational approach with the previously published decay rates, we find an overall good agreement. This is illustrated in Figures 2--5, where our calculations are compared with the experimental data reported by Salih {\it et al} (1985), Crespo L\'opez-Urrutia {\it et al} (1994), Mullman {\it et al} (1998) and the those obtained theoretically by Raassen {\it et al} (1998). More precisely, as shown in Figure 2, we obtain a mean ratio of 1.036 $\pm$ 0.205 when comparing our transition probabilities with those of Salih {\it et al} (1985), who combined lifetime values obtained by TR-LIF with branching fractions measured on spectra recorded with the 1-m Fourier-transform spectrometer at the Kitt Peak National Observatory, to deduce the decay rates for 41 transitions depopulating the 3d$^7$4p z$^5$F, z$^5$D and z$^5$G levels. Figure 3 shows a slighly larger scatter between our calculations and the transition probabilities published by Crespo L\'opez-Urrutia {\it et al} (1994). Here the mean ratio $gA_{\textup{This work}}$/$gA_{\textup{Crespo}}$ is equal to 1.098 $\pm$ 0.493. However, it is worth noting that the $gA$-values obtained by the latter authors were deduced from the combination of branching ratio measurements with available experimental lifetimes for the 3d$^7$4p z$^5$F, z$^5$D, z$^5$G levels Pinnington {\it et al} 1973, 1974, Salih {\it et al} 1985) but also with estimated lifetimes for the z$^3$G, z$^3$F and z$^3$D levels from the measurement of total intensity of all lines of each of these levels under the assumption of almost equal population. The branching fractions were found using intensity measurements with a special hollow electrode r.f. discharge and using a phase method with a modified wall-stabilized arc in a spectro-interferometric arrangement. Mullman {\it et al} (1998) reported 28 oscillator strengths for ultraviolet \ion{Co}{ii} lines deduced from laser-induced fluorescence lifetimes and branching fraction measurements using a high resolution grating spectrometer and an optically thin hollow cathode discharge. As illustrated in Figure 4, our $gf$-values were found to agree in general within 20--30 \% with the results obtained by Mullman {\it et al} (1998). More particularly, the mean ratio $gf_{\textup{This work}}$/$gf_{\textup{Mullman}}$ was found to be equal to 1.339 $\pm$ 0.552 when using all the common lines, and to 1.238 $\pm$ 0.411 when using the strongest transitions for which log $gf$ $>$ -1. Finally, Figure 5 shows the comparison between our computed oscillator strengths and those published by Raassen {\it et al} (1998) who used the theoretical method of orthogonal operators to determine the log $gf$-values for (3d$^8$ + 3d$^7$4s) -- 3d$^7$4p transitions in \ion{Co}{ii}. In this case, the two sets of data generally agree within about 20\%, this percentage difference being reduced to 12\% when considering the most intense transitions with log $gf$ $>$ -1. We also note that our oscillator strengths are on average larger than those obtained by Raassen {\it et al}. This could be due to the fact that these latter authors included explicitly a less extended set of interacting configurations in their model.

   \begin{figure*}
   \includegraphics[width=15cm,clip]{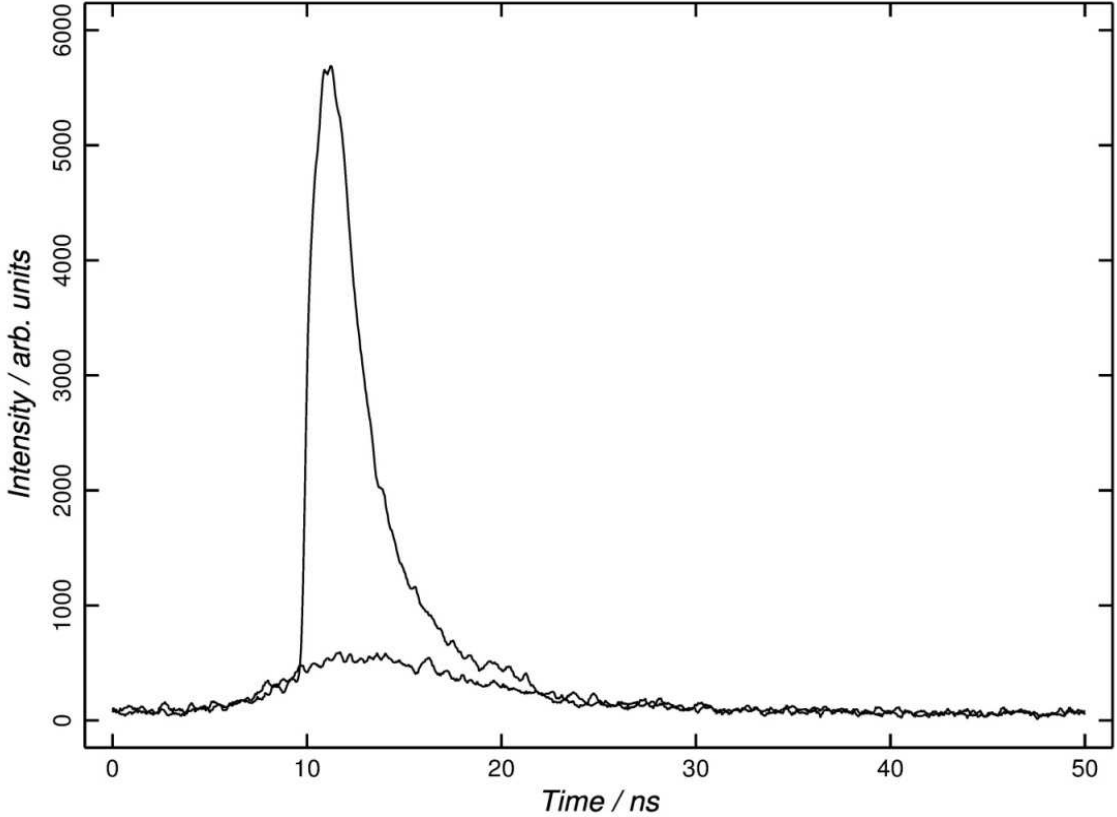}
      \caption{Decay of the 4d$^7$($^4$F)4d e$^5$H$_6$ level at 236 nm perturbed by the fluorescence from the 4d$^7$($^4$F)4p z$^5$G$_5$ level at 231 nm excited by the first-step laser. The lower curve shows a measurement at 236 nm with the second-step laser turned off, revealing the contribution from z$^5$G$_5$.}
         \label{Fig1}
   \end{figure*}

  \begin{figure*}
   \includegraphics[width=15cm,clip]{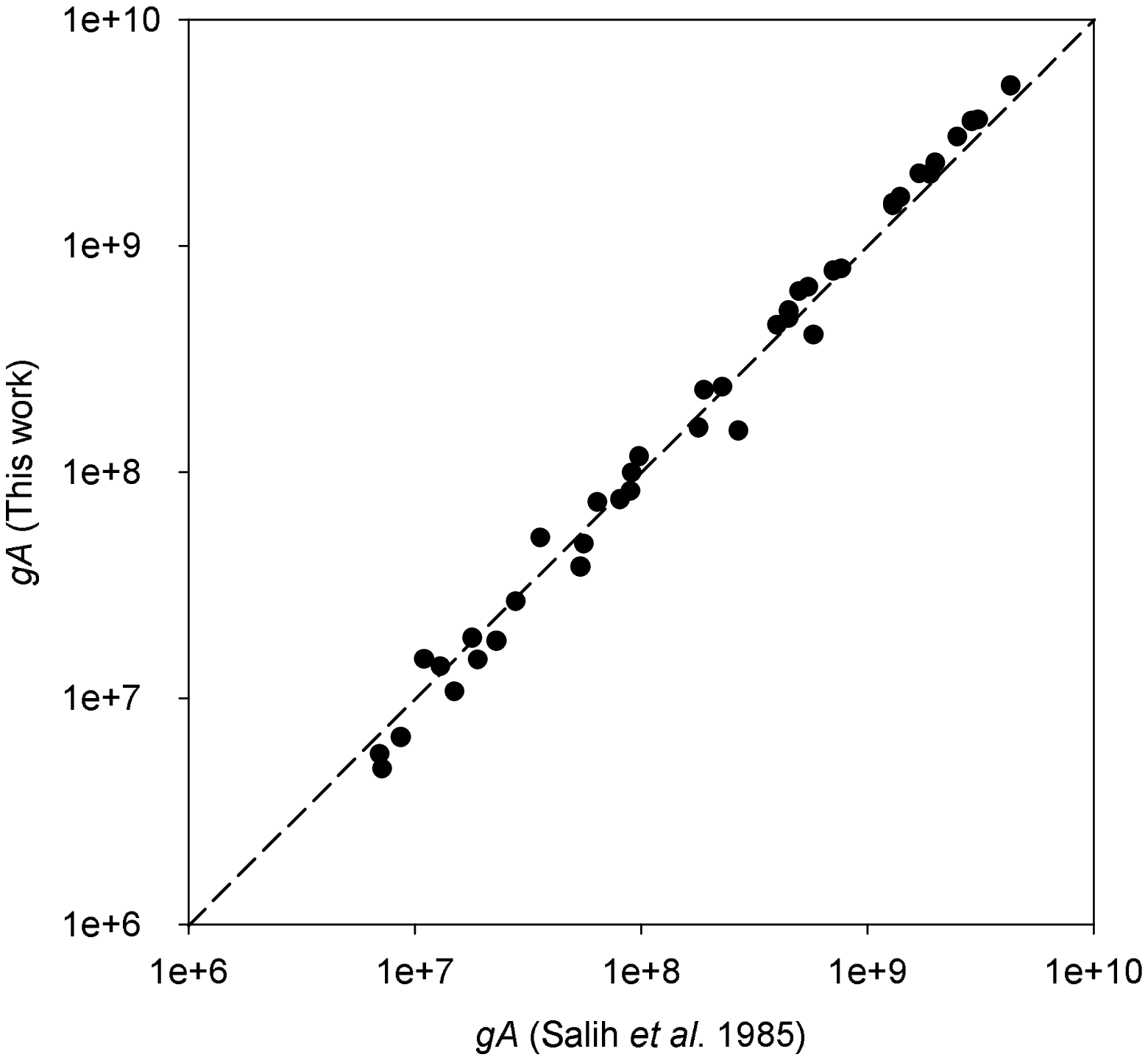}
      \caption{Comparison between the transition probabilities ($gA$ in s$^{-1}$) calculated in the present work and those deduced from the experimental measurements due to Salih {\it et al} (1985).}
         \label{Fig2}
   \end{figure*}

      \begin{figure*}
   \includegraphics[width=15cm,clip]{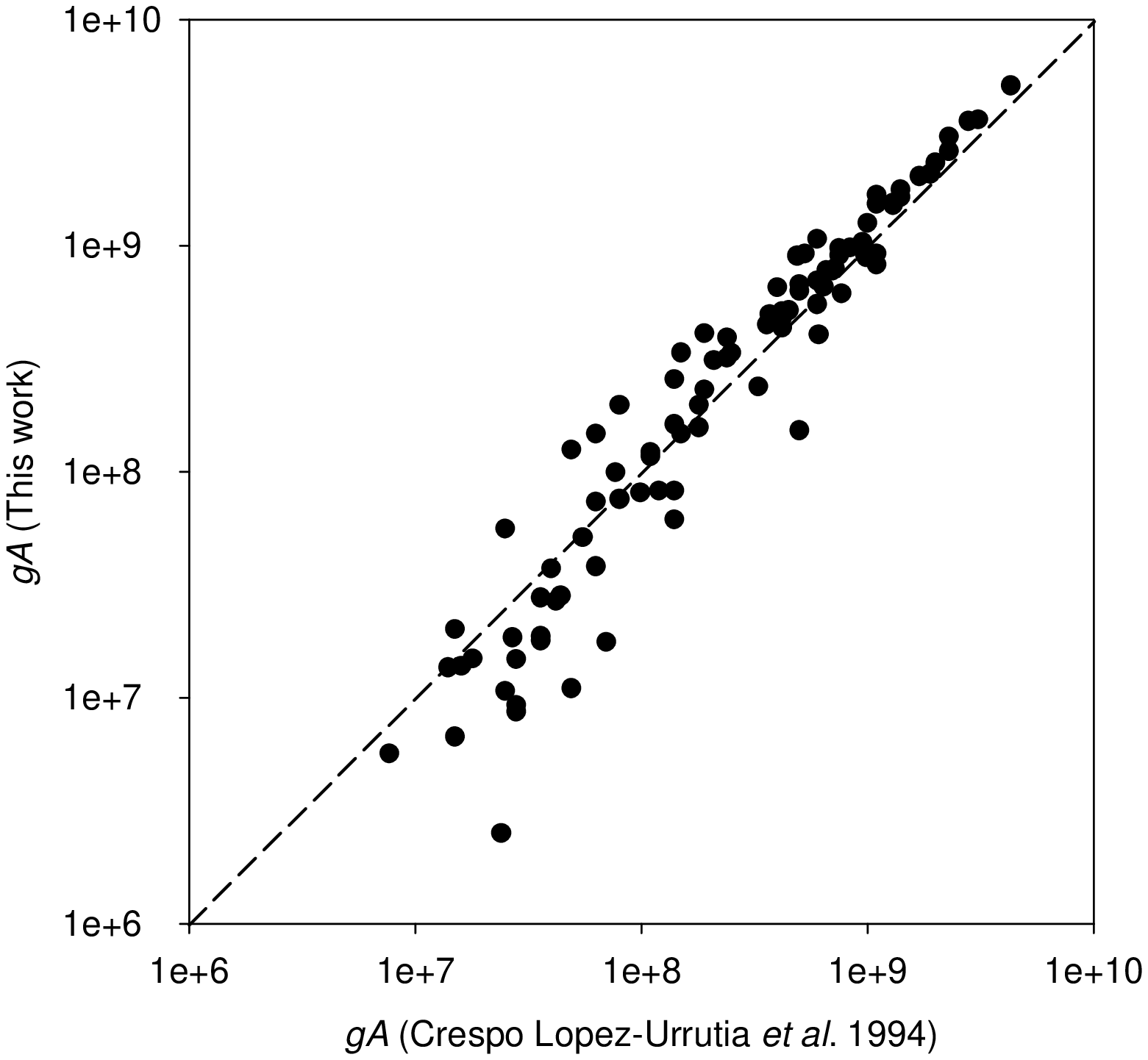}
      \caption{Comparison between the transition probabilities ($gA$ in s$^{-1}$) calculated in the present work and those deduced from the experimental measurements due to Crespo Lopez-Urrutia {\it et al} (1994).}
        \label{Fig3}
   \end{figure*}

      \begin{figure*}
   \includegraphics[width=15cm,clip]{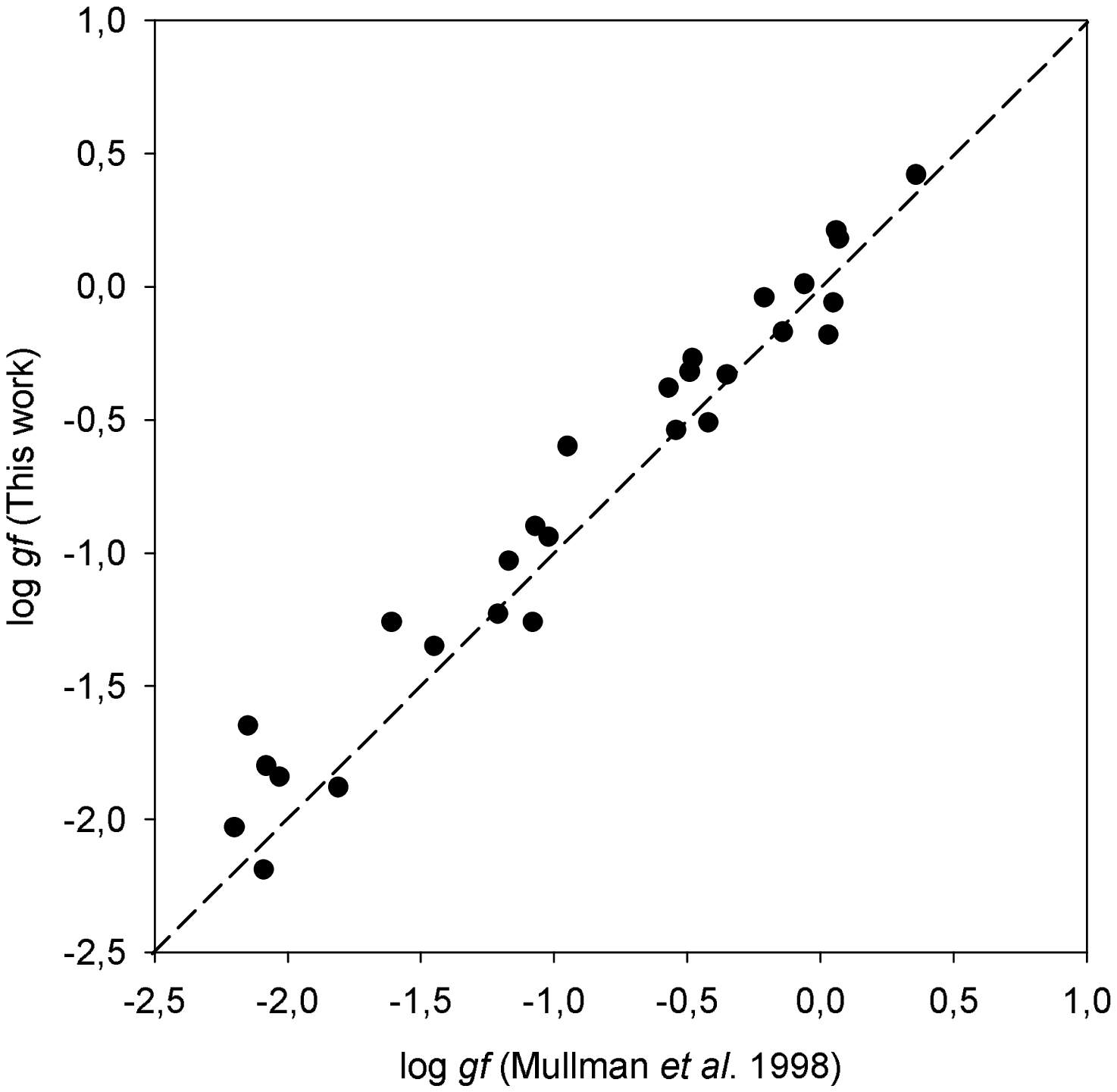}
      \caption{Comparison between the oscillator strengths (log $gf$) calculated in the present work and those deduced from the experimental measurements due to Mullman {\it et al} (1998).}
         \label{Fig1}
   \end{figure*}

         \begin{figure*}
   \includegraphics[width=15cm,clip]{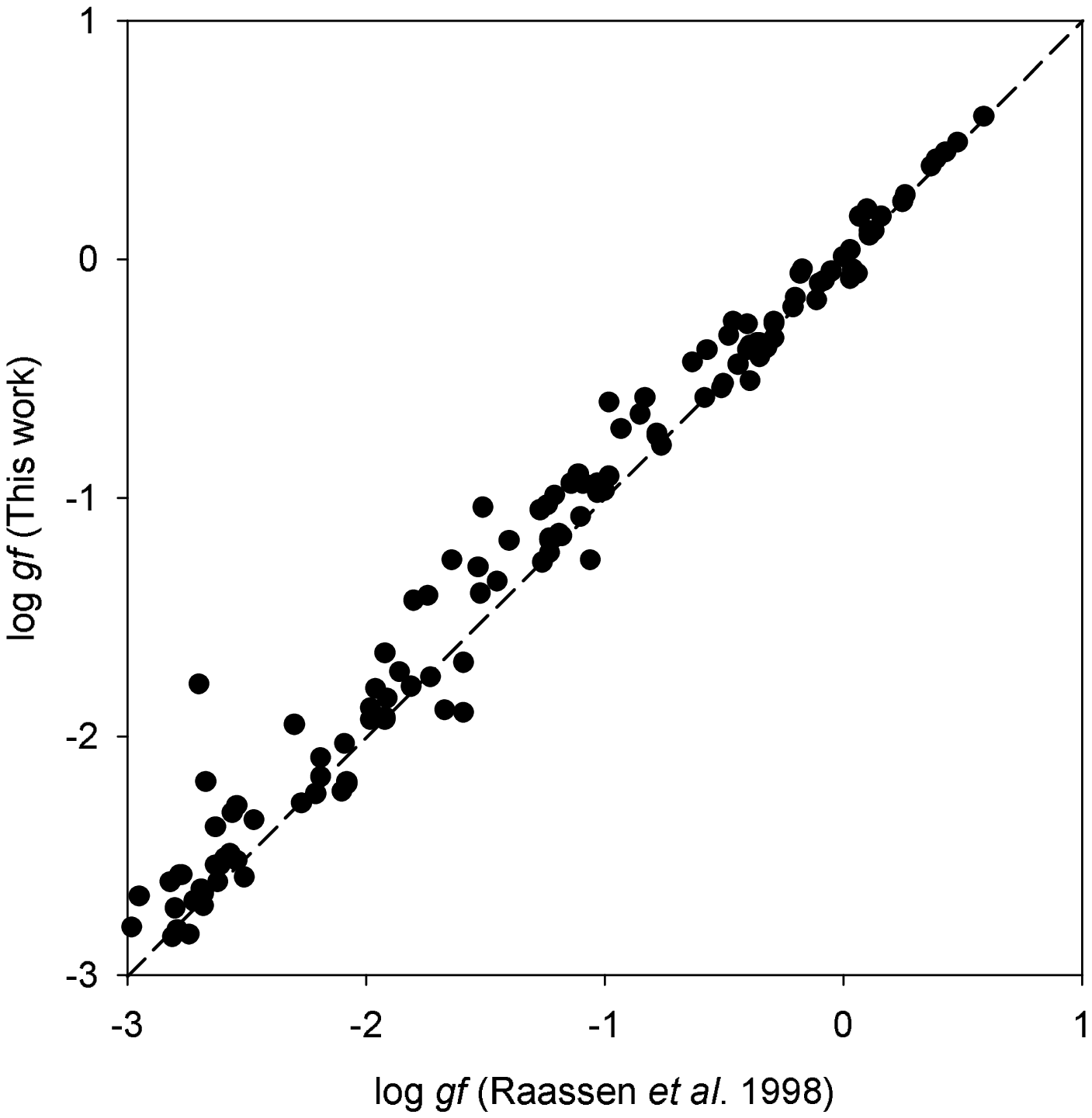}
      \caption{Comparison between the oscillator strengths (log $gf$) calculated in the present work and those computed by Raassen {\it et al} (1998).}
         \label{Fig1}
   \end{figure*}

\section{Conclusion}

Transition probabilities and oscillator strengths have been obtained for 5080 spectral lines in \ion{Co}{ii} using the pseudo-relativistic Hartree-Fock method including the most important intravalence correlation and core-polarization effects. The accuracy of the new data has been assessed through detailed comparisons with previously published experimental and theoretical radiative rates together with new lifetime measurements performed in the present work using the laser-induced fluorescence technique with one- and two-step excitations. In view of the overall agreement obtained between all sets of results, it is expected that the new $gA$- and $gf$-values should be accurate to a few percent for the strongest transitions and to within 20--30\% for weaker lines.

\section*{Acknowledgements}

This work was financially supported by the Integrated Initiative of Infrastructure Project LASERLAB-EUROPE, contract LLC002130, by the Belgian F.R.S.-FNRS, and by the Swedish Research Council through the Linnaeus grant to the Lund Laser Centre and the Knut and Alice Wallenberg Foundation. P.Q. and P.P. are, respectively, Research Director and Research Associate of the F.R.S.-FNRS., V.F. is currently a post-doctoral researcher of the Return Grant program of the Belgian Scientific Policy (BELSPO) and H.H. gratefully acknowledges the grant no 621-2011-4206 from the Swedish Research Council. The Belgian team is grateful to the Swedish colleagues for the warm hospitality enjoyed at the Lund Laser Centre during the two campaigns in June and August 2015.

\nocite{a}
\nocite{b}
\nocite{c}
\nocite{d}
\nocite{e}
\nocite{f}
\nocite{g}
\nocite{h}
\nocite{i}
\nocite{j}
\nocite{k}
\nocite{l}
\nocite{m}
\nocite{n}
\nocite{o}
\nocite{p}
\nocite{q}
\nocite{r}
\nocite{s}
\nocite{t}
\nocite{u}
\nocite{v}
\nocite{x}
\nocite{y}
\nocite{z}
\nocite{aa}
\nocite{ab}




\begin{thebibliography}{99}

\bibitem[\protect\citeauthoryear{a}{2015}]{a}
Battistini C. \& Bensby T., 2015, Astron. Astrophys., 577, A9

\bibitem[\protect\citeauthoryear{b}{2012}]{b}
Bravo E. \& Martinez-Pinedo G., 2012, Phys. Rev. C, 85, 055805

\bibitem[\protect\citeauthoryear{c}{1981}]{c}
Cowan R.D., 1981, The Theory of Atomic Structure and Spectra (California University Press,
Berkeley)

\bibitem[\protect\citeauthoryear{d}{2012}]{d}
Crespo L\'opez-Urrutia J.R., Ulbel M., Neger T. \& J\"ager H., 1994, J. Quant. Spectrosc. Radiat. Transfer, 52, 89

\bibitem[\protect\citeauthoryear{e}{2012}]{e}
Engstr\"om L., Lundberg H., Nilsson H., Hartman H. \& B\"ackstr\"om E., 2014, Astron. Astrophys., 570, A34

\bibitem[\protect\citeauthoryear{f}{2012}]{f}
Fraga S., Karwowski J., Saxena K.M.S., 1976, Handbook of Atomic Data (Elsevier, Amsterdam)

\bibitem[\protect\citeauthoryear{g}{2012}]{g}
Iglesias L., 1979, Opt. Pura Apl., 12, 63

\bibitem[\protect\citeauthoryear{h}{2012}]{h}
Kramida A., Ralchenko Yu., Reader J. \& NIST ASD Team, 2014, NIST Atomic Spectra Database, Available : http://physics.nist.gov/asd

\bibitem[\protect\citeauthoryear{i}{2012}]{i}
Kurucz R., 2011, http://kurucz.harvard.edu/atoms/2801/gf2801.life

\bibitem[\protect\citeauthoryear{j}{2012}]{j}
Lawler J.E., Sneden C. \& Cowan J.J., 2015, Astrophys. J. Suppl. Ser., 220, 13

\bibitem[\protect\citeauthoryear{k}{2012}]{k}
Lind K., Bergemann M. \& Asplund M., 2012, Mon. Not. R. Astron. Soc., 427, 50

\bibitem[\protect\citeauthoryear{l}{2012}]{l}
Lundberg H., Hartman H., Engstr\"om L., Nilsson H., Persson A., Palmeri P., Quinet P., Fivet V., Malcheva G. \& Blagoev K., 2016, Mon. Not. R. Astron. Soc. (in press) [DOI 10.1093/mnras/stw922]

\bibitem[\protect\citeauthoryear{m}{2012}]{m}
Mullman K.L., Cooper J.C. \& Lawler J.E., 1998, Astrophys. J., 495, 503

\bibitem[\protect\citeauthoryear{n}{2012}]{n}
Mullman K.L., Lawler J.E., Zsarg\'o J. \& Federman S.R., 1998, Astrophys. J., 500, 1064

\bibitem[\protect\citeauthoryear{o}{2012}]{o}
Nave G. \& Sansonetti C.J., 2011, J. Opt. Soc. Am. B, 28, 737

\bibitem[\protect\citeauthoryear{p}{2012}]{p}
Pagel B.E.J., 2009, Nucleosynthesis and Chemical Evolution of Galaxies (Cambridge University Press)

\bibitem[\protect\citeauthoryear{q}{2012}]{q}
Palmeri P., Quinet P., Fivet V., Bi\'emont E., Nilsson H., Engstr\"om L. \& Lundberg H., 2008, Phys. Scr., 78, 015304

\bibitem[\protect\citeauthoryear{r}{2012}]{r}
Pickering J.C., 1998, Phys. Scr., 58, 457

\bibitem[\protect\citeauthoryear{s}{2012}]{s}
Pickering J.C., Raassen A.J.J., Uylings P.H.M. \& Johansson S., 1998, Astrophys. J. Suppl. Ser., 117, 261

\bibitem[\protect\citeauthoryear{t}{2012}]{t}
Pinnington E.H., Lutz H.O. \& Carriveau G.W., 1973, Nucl. Instrum. Meth., 110, 55

\bibitem[\protect\citeauthoryear{u}{2012}]{u}
Pinnington E.H., Lutz H.O. \& Carriveau G.W., 1974, Z. Phys., 267, 27

\bibitem[\protect\citeauthoryear{v}{2012}]{v}
Quinet P., Palmeri P., Bi\'emont E., Li Z.S., Zhang Z.G. \& Svanberg S., 2002, J. Alloys Comp., 344, 255

\bibitem[\protect\citeauthoryear{x}{2012}]{x}
Quinet P., Palmeri P., Bi\'emont E., McCurdy M.M., Rieger G., Pinnington E.H.,
Wickliffe M.E., Lawler J.E., 1999, Mon. Not. R. Astron. Soc., 307, 934

\bibitem[\protect\citeauthoryear{y}{2012}]{y}
Raassen A.J.J., Pickering J.C. \& Uylings P.H.M., 1998, Astron. Astrophys. Suppl. Ser., 130, 541

\bibitem[\protect\citeauthoryear{z}{2012}]{z}
Salih S., Lawler J.E. \& Whaling W., 1985, Phys. Rev. A, 31, 744

\bibitem[\protect\citeauthoryear{aa}{2012}]{aa}
S\o rensen G., J. Phys. (Paris) Colloques, 1979, 40, C1-190

\bibitem[\protect\citeauthoryear{ab}{2012}]{ab}
Woosley S.E. \& Weaver T.A., 1995, Astrophys. J. Suppl. Ser., 101, 181

\end{thebibliography}




%
%


\bsp	
\label{lastpage}
\end{document}